\documentclass[runningheads]{svmult}

\usepackage{makeidx}   
\usepackage{graphicx}  
\usepackage{subeqnar}  
\usepackage{multicol}  
\usepackage{physprbb}  
\makeindex             

\def\ltsim{~\rlap{$<$}{\lower 1.0ex\hbox{$\sim$}}~}
\def\gtsim{~\rlap{$>$}{\lower 1.0ex\hbox{$\sim$}}~}
\newcommand{\HI}{\mbox {\sc H\thinspace{i}}}
\newcommand{\HII}{\mbox {\sc H\thinspace{ii}}}
\newcommand{\Hline}[1]{\mbox{H{\footnotesize {#1}}}}
\newcommand{\Halpha}{\Hline{\mbox{$\alpha$}}}
\newcommand{\kms}{\mbox{km\thinspace s$^{-1}$}}

\begin{document}
\title*{Starbursts, Dark Matter And Dwarf Galaxy Evolution}
\toctitle{Starbursts, Dark Matter And Dwarf Galaxy Evolution}

\author{Gerhardt R. Meurer}
\authorrunning{Gerhardt R. Meurer}

\institute{The Johns Hopkins University, Baltimore MD 21218, USA}

\maketitle

\begin{abstract}
Optical and \HI\ imaging of both dwarf irregular (dI) and Blue Compact
Dwarf (BCD) galaxies reveal important clues on how dwarf galaxies
evolve and their star formation is regulated.  Both usually show
evidence for stellar and gaseous disks.  However, their total mass is
dominated by dark matter.  Gas rich dwarfs form with a range of disk
structural properties.  These have been arbitrarily separated them into
two classes on the basis of central surface brightness.  Dwarfs with
$\mu_{0}(B) \ltsim 22\, {\rm mag\, arcsec^{-2}}$ are usually classified
as BCDs, while those fainter than limit are usually classified as
dIs.  Both classes experience bursts of star formation, but with an
absolute intensity correlated with the disk surface brightness.  Even in
BCDs the bursts typically represent only a modest \ltsim{1} mag
enhancement to the $B$ luminosity of the disk.  While starbursts are
observed to power significant galactic winds, the fractional ISM loss
remains modest.  Dark matter halos play an important role in determining
dwarf galaxy morphology by setting the equilibrium surface brightness of
the disk.
\end{abstract}

\section{Introduction}

Despite their morphological differences dwarf irregular (dI), Blue
Compact Dwarf (BCD), and dwarf elliptical (dE) galaxies have similar
optical structures - their radial profiles are exponential, at least
at large radii (e.g.\ \cite{bmcm86,cb87,mmhs97}).  Are there
evolutionary connections between these morphologies?  One scenario
expounded by Davies \&\ Phillips \cite{dp88} starts with an initial dI
galaxy; if its ISM manages to concentrate at the center of the galaxy
a tremendous starburst occurs resulting in a BCD morphology.  This
starburst powers a galactic wind (e.g.\ \cite{ds86,mhws95}).  If the
wind is strong enough all of the ISM is expelled resulting in a dE
morphology.  If some ISM remains, the system fades back into a dI, and
undergoes a few more dI $\Leftrightarrow$ BCD transitions before
eventually expelling all of its ISM to become a dE. Here I will
address the validity of this scenario.  In Sec.~\ref{s:opt}, I compare
the optical structure of dIs and BCDs; Sec.~\ref{s:rad} details the
\HI\ structure and dynamics of two BCDs: NGC~1705 ($D = 6.2$ Mpc) and
NGC~2915 ($D = 3.1$ Mpc), and compares them to dI galaxies; and
Sec.~\ref{s:syn} synthesizes the optical and radio results to form a
new scenario where Dark Matter (DM) plays a dominant role in
determining the morphology of gas rich dwarfs.  Here I adopt $H_0 =
75\, {\rm km\, s^{-1}\, Mpc^{-1}}$.

\section{The optical structure and classification of gas rich dwarfs\label{s:opt}}

The exponential profile portion of BCDs and dIs probably signifies
the presence of a rotating disk, which are certainly typical of the gas
distributions in these systems \cite{v98,dm97}.  In addition to the
disk component, BCDs usually have a blue high surface brightness
excess of light in their centers, rich in \HII\ emission and young
star clusters \cite{m95,mmhs97}.  I will call this blue excess the starburst,
since it is responsible for the starburst characteristics of BCDs.
The integrated colors of this component, after subtracting off the
disk, indicate that it must be due to a young stellar population with
an age $\sim 10$ Myr if instantaneous burst models are adopted or
$\sim 100$ Myr if constant star formation rate models are adopted
\cite{mmh99}.  The strong \Halpha\ fluxes are more consistent with the
constant star formation rate models (because $\sim 10$ Myr old
instantaneous bursts are no longer ionizing).  In either case this is
much less than the Hubble time, confirming the starburst nature
of this component.  In comparison, the colors of the disk are
typically like those of stellar populations forming continuously over
a Hubble time (i.e.\ like dI galaxies, cf. \cite{pt96}), or a bit
redder suggesting an inactive population. 

Surface brightness profile fitting provides a means to determine both
the relative strength of the starburst, and the structural properties
of the disk (see \cite{mmh99} for details).  The outer portions of the
profile are fitted with an exponential, yielding the extrapolated
central surface brightness $\mu_0$, and scale length $\alpha^{-1}$ of
the disk.  The burst and disk are separated by assuming that the disk
remains exponential all the way into the center.  The strongest
starbursts are about twice as bright as their hosts.  Hence, while
starbursts can outshine the host disk they are nevertheless
modest\ltsim{1} mag enhancements to the total $B$ flux of
BCDs. Typical flux enhancements are only a few tens of percent.  About
20\% of BCDs have exponential profiles all the way into their cores,
hence they show no structural evidence for a starburst.  The mass
contribution of the starburst is even smaller than the flux
contribution, typically \ltsim{5}\%.  These are not the $\sim 6$ mag
starburst enhancements proposed to explain the excess of faint blue
galaxies at moderate redshifts \cite{bf96}.

\begin{figure}
\begin{center}
\includegraphics[width=0.48\textwidth]{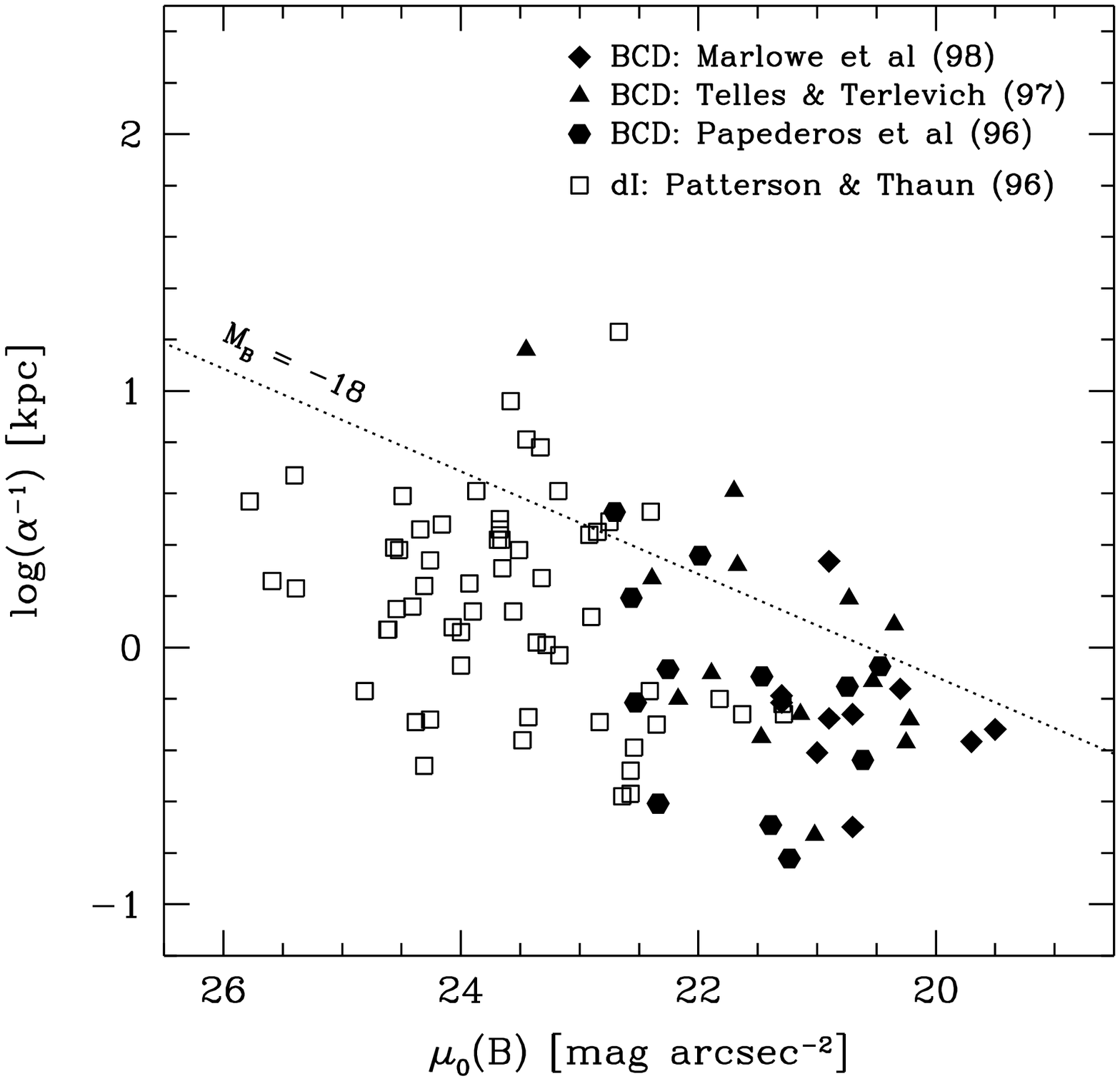}
\includegraphics[width=0.48\textwidth]{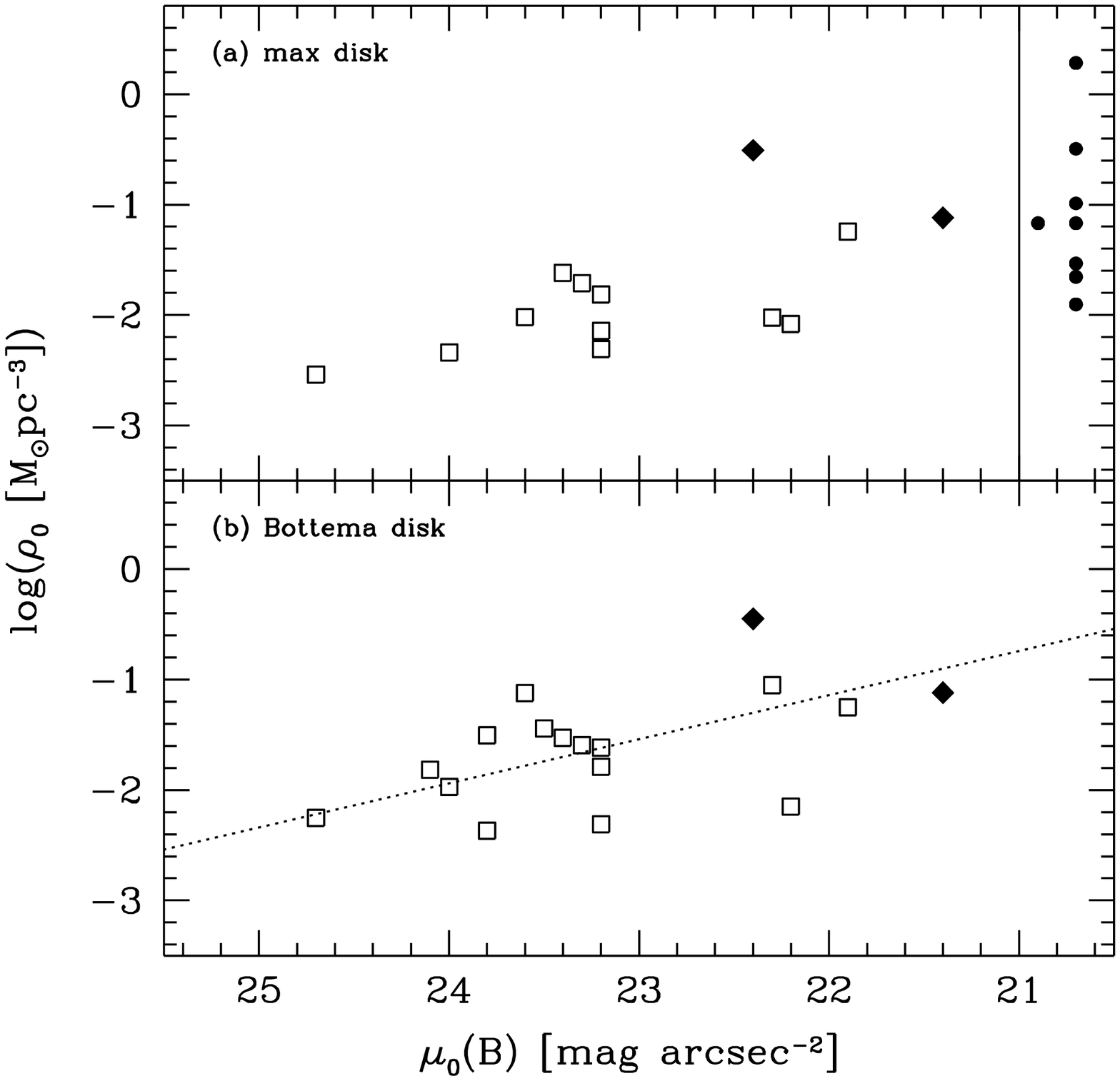}
\end{center}
\caption{(Left) Exponential disk parameters for samples of BCD and dI
galaxies \cite{mmh99,tt97,pltf96,pt96}. \label{f:struct}}
\end{figure}
\begin{figure}
\vspace{-0.5cm}
\caption{(Right) DM halo central density $\rho_0$ plotted against disk
central surface brightness.  Open squares are from de Blok \&\ McGaugh
\cite{dm97}, while diamonds represent NGC~1705 and NGC~2915.  The top
panel shows the results for maximimum disk model fits, while the bottom
panel shows Bottema disk fits.  The circles on the right side of the top
panel mark crude estimates of $\rho_0$ in 12 BCDs with published and
unpublished RCs.  The dotted line, at bottom,  is a fit to the Bottema disk
results with the relationship $\log(\rho_0) = 0.4 \log(\mu_0) + {\rm
Constant}$.  \label{f:rho0}}
\end{figure}

Figure~\ref{f:struct} compares the disk parameters $\mu_0$, 
$\alpha^{-1}$ of both BCDs and dI galaxies.  Note, $\mu_0$ does not
include the contribution of the starburst core.  While there is some
overlap, we see that BCD distribution is offset from that of dIs
particularly in $\mu_0$ which typically is 2.5 mag arcsec$^{-2}$ more
intense in BCDs than in dIs.  Structurally BCD disks are very different
from those of dIs.

The absence of BCDs on the left half of Fig.~\ref{f:struct} is
puzzling.  Does this mean that dI galaxies do not experience
starbursts?  Examination of surface brightness and color profiles
\cite{pt96} reveal several dIs with an exponential profile, and a higher
surface brightness blue excess, structural evidence for starbursts in
dIs.  The episodic star-forming nature of dIs is well demonstrated
using color-magnitude diagrams of the nearest ones (e.g.\
\cite{gtdschsm98}). However the observed central surface brightness of
the bursting dIs {\em including the light of this central excess} is
typically $\mu(B) \gtsim 22\, {\rm mag\, arcsec^{-1}}$, much fainter
than the central regions of BCDs.  While dIs do experience starbursts,
they are pathetic and not usually recognized as such since they are
not {\em intense\/} enough.  We see that there are both BCDs and dIs
with central starbursts, as well as cases of both types that are
exponential all the way into their centers.  The separation in to two
classes appears to be an arbitrary segregation by central surface
brightness of the underlying disk at $\mu_0(B) \approx 22$ mag
arcsec$^{-2}$.

\section{\HI\ structure and dynamics of BCDs\label{s:rad}}

Compared to dIs there are not that many \HI\ imaging studies of BCDs;
in part due to their small numbers and usual compact angular sizes.
Two that have been well imaged in \HI\ are NGC~2915 \cite{mcbf96}, and
NGC~1705 \cite{msk98}.

Both galaxies show evidence of star formation churning up the neutral
ISM.  In NGC~2915, enhancements in the velocity dispersion correspond
well to \Halpha\ bubbles and peculiar star formation knots.  However
it does not appear that \HI\ is being ejected from the system.  In the
center of the galaxy, where star formation is the most vigorous,
$\sigma_{\rm HI} \approx 40$ \kms\ which is the same as the one
dimensional velocity dispersion derived for the DM particles.  Hence,
star formation appears to be maintaining the central \HI\ in virial
equilibrium with the DM halo.  This suggests that DM plays a role in
the feedback process: if the starburst energizes \HI\ to have
$\sigma_{\rm HI}$ much larger than the halo velocity dispersion, then
neutral ISM is thrown into the halo (or beyond) halting star formation. 

NGC~1705 displays a strong galactic wind in \Halpha.  There is a spur
of \HI\ emission obliquely jutting out from its \HI\ disk that appears
to be a neutral ISM extension of this wind.  If so, then NGC~1705 has
ejected about 8\%\ of its neutral ISM at a mass loss rate at least
comparable to the star formation rate.  Nevertheless, even in this BCD
with one of the most spectacular \Halpha\ outflows, the majority of
the ISM is retained in a disk.  This starburst is incapable of totally
blowing away the ISM.

Although both galaxies have some kinematic irregularities their dominant
structures are extended rotating disks which are strongly centrally
peaked.  These are typical properties of BCDs imaged in \HI\
\cite{tbps94,v98}.  The disk of NGC~2915 is so extended that it has
the \HI\ appearance of a late type barred spiral.  Similar galaxies
include IC~10 \cite{w98} and NGC~4449 \cite{h98}.  

The rotation curves of both NGC~1705 and NGC~2915 show a fairly steep
rise over the optical face of the galaxy which then becomes flat out
to the edge of the \HI\ distribution.  They are the first BCDs with
mass model fits to their rotation curves.  The mass models include (1)
a stellar distribution whose mass to light ratio is given by either a
maximum disk model or by the optical colors (the ``Bottema disk''
model \cite{dm97}); (2) the neutral ISM distribution; and (3) a dark
matter halo.  This halo is taken to be a pseudo-isothermal sphere
whose free parameters are the central density $\rho_0$ and the core
radius $R_c$. From these, the asymptotic rotational velocity
and halo velocity dispersion can be calculated \cite{lsv90}.
In both galaxies, DM dominates the mass distribution, even within the
optical radius of the galaxy.  In comparison, the stellar component
has a mass equal to or {\em less than\/} the neutral gas disk.

Overall, the global dynamics of BCDs appear to be similar to dIs: they
are dominated by rotating disks with normal looking RCs.  A
distinction between the two types is seen when the DM halo densities
$\rho_0$ are compared, as shown in Fig.~\ref{f:rho0}.  Central
densities found by maximum disk and Bottema disk fits are shown in
separate panels.  The comparison sample is taken from de Blok \&\
McGaugh \cite{dm97}, and includes only galaxies with $M_B > -18$ mag.
This comparison shows that NGC~1705 and NGC~2915 have two of the
highest $\rho_0$ measurements of any dwarf galaxies.  In order to
check that these galaxies are typical, I crudely estimated $\rho_0$
from the central velocity gradient for 12 BCDs with published or
unpublished RCs, and plotted them as circles at arbitrary $\mu_0$ in
the top panel of Fig.~\ref{f:rho0}.  These estimates are upper limits,
since the contribution of the baryonic components to the velocity
gradients have not been removed.  Nevertheless, the comparison
indicates that NGC~1705 and NGC~2915 have normal $\rho_0$ for BCDs.
Figure~\ref{f:rho0} shows a weak but noticeable correlation between
$\log(\rho_0)$ and $\mu_0(B)$, with higher surface brightness disks
corresponding to higher $\rho_0$ halos.  This result was first noted
in dIs by de Blok \&\ McGaugh \cite{dm97}.

\section{Evolutionary Connections\label{s:syn}}

The correlation in Fig.~\ref{f:rho0} can readily be explained by
considering the response of a self gravitating disk immersed in a
dominant DM halo core of constant density $\rho_0$, i.e.\ where the
rotation curve is linearly rising.  If the disk is maintained at
constant stability parameter and the star formation rate per unit area
scales with the gas density divided by the dynamical time \cite{k98},
then it is straight-forward to show that the surface brightness should
scale with $\rho_0$ \cite{msk98,moriond}. This is consistent with the
observed correlation, as shown by the dotted line in
Fig.~\ref{f:rho0}.  A similar correlation between surface brightness
and $\rho_0$ holds in the center of larger starburst galaxies
\cite{mhlll97}.  However for them it is normal baryonic matter that
dominates $\rho_0$ rather than DM. In essence, the central mass
density determines the equilibrium star formation rate of the embedded
disk.  

Following from the discussion in Sec.~ref{s:opt}, the disk central
intensity largely determines whether a dwarf galaxy is classified as a
BCD or dI.  The optical size of dwarfs seems to be limited to the core
radius $R_c$.  Hence, both DM halo parameters are important in governing
the morphology of gas rich dwarfs.

Can there be evolution between dI and BCD classes?  While some evolution
in $\rho_0$ may be allowed, it is unlikely that there can be enough to
change a typical dI into a typical BCD. That would require a 2.5 mag
arcsec$^{-2}$ change in $\mu_0$, or equivalently, a factor of ten change
in $\rho_0$.  To do this with a mass loss or gain would require a 55\%\
change in mass if the expansion or contraction is homologous
\cite{moriond}. The problem is that there isn't that much {\em
baryonic\/} mass in a dwarf galaxy.  To effect this large of a change
would require DM loss or gain.  This is not feasible if DM is
non-dissipative and feels only the force of gravity, as is usually
assumed.  I conclude that there is probably little dI $\Leftrightarrow$
BCD evolution.

If the ISM were removed from a dI or BCD, it could still plausibly
evolve into a dE galaxy.  However, as noted in
\S~\ref{s:rad} even in a dwarf galaxy undergoing a strong starburst with
a spectacular galactic wind (NGC~1705), the fractional loss of the ISM
is modest.  If this is typical, it would take on order of 10 bursts to
expel all the ISM from a BCD.  The bursts aren't strong enough, and the
ISM distributions are too flattened to allow a single burst expulsion of
the ISM \cite{dh94}.  The demographics of dwarf galaxy morphologies
point to an environmental component to their evolution.  Gas rich dwarfs
are found in low density environments where the frequency of external
starburst triggers is low.  They survive easily. The clock runs faster
(more frequent triggers) in clusters, and in addition ram pressure
stripping would accelerate the removal of gas from dwarfs, while tidal
truncation of DM halos would assist galactic wind losses. Hence it is
not surprising that gas poor dEs are found more often in clusters than
the field.

\section{Conclusions}

We are now at a position to re-evaluate the Davies and Phillips
\cite{dp88} scenario for dwarf galaxy evolution.  The mechanisms they
invoke have clearly been verified.  Dwarf galaxies do experience
starbursts and these can expel some of the ISM.  Mass expulsion can
rival or surpass lock up into stars in regulating the gas content of
dwarfs.  However, the results of any single burst are not so severe.
Cataclysmic bursts are not common at the present epoch, and the milder
bursts that are observed may not be sufficient to change a galaxy's
morphological classification.  The morphology of a dwarf galaxy is
largely set by its enveloping dark halo, and is relatively impervious to
starbursts.

{\bf Acknowledgements}: I thank the organizers for asking me to talk on
this subject.  Since this was done as a last minute replacement for
Daniel Kunth, I was not able to prepare anything new.  Instead this talk
is based on my 1998 Moriond presentation \cite{moriond}.  I thank my
collaborators on the original projects this work is based on: Sylvie
Beaulieu, Claude Carignan, Ken Freeman,
Tim Heckman, Neil Killeen, Amanda Marlowe, and Lister
Staveley-Smith.

\end{document}